\documentclass[twocolumn,aps,superscriptaddress,nofootinbib,floatfix]{revtex4}
\usepackage{float}
\usepackage{epsfig,bm,feynmf}
\usepackage{graphics}

\usepackage[normalem]{ulem}  
\usepackage[dvips]{color} 

\renewcommand\sout{\bgroup \color{red} \ULdepth=-.5ex \ULset}

\begin{document}
	
	\title{Spinodal instabilities of baryon-rich quark matter in heavy ion collisions}
	\author{Feng Li}\email{lifengphysics@gamil.com}
	\affiliation{Cyclotron Institute and Department of Physics and
		Astronomy, Texas A$\&$M University, College Station, TX 77843-3366,
		USA}
	\author{Che Ming Ko}\email{ko@comp.tamu.edu}
	\affiliation{Cyclotron Institute and Department of Physics and
		Astronomy, Texas A$\&$M University, College Station, TX 77843-3366, USA}
	
\begin{abstract}
Using the test-particle method to solve the transport equation derived from the Nambu-Jona-Lasino (NJL) model, we study how phase separation occurs in an expanding quark matter like that in a heavy ion collision. To test our method, we first investigate the growth rates of unstable modes of quark matter in a static cubic box and find them to agree with the analytical results that were previously obtained using the linear response theory.  In this case, we also study the higher-order scaled density moments in the quark matter, which have values of one for a uniform density distribution, and they are found to increase with time and saturate at values significantly larger than one after the phase separation.   The skewness of the quark number event-by-event distribution in a small sub-volume of the system is also found to increase, but this feature disappears if the sub-volume is large.  For the expanding quark matter, two cases are considered with one using a blast-wave model for the initial conditions and the other using initial conditions from a mulple-phase transport (AMPT) model. In both cases, we find the expansion of the quark matter is slowed down by the presence of a first-order phase transition.  Also, density clumps appear in the system and the momentum distribution of partons becomes anisotropic , which can be characterized by large  scaled density moments and non-vanishing anisotropic elliptic and quadrupolar flows, respectively.  The large density fluctuations further lead to an enhancement in the dilepton yield.   In the case with the AMPT initial conditions, the presence of a first-order phase transition also results in  a narrower rapidity distribution of partons after their freeze out.   These effects of density fluctuations can be regarded as possible signals for a first-order phase transition that occurs in the baryon-rich quark matter formed in relativistic heavy ion collisions.

\end{abstract}

\maketitle
\section{Introduction}

Studying the properties of baryon-rich quark-gluon plasma (QGP) is the main focus of  the beam energy scan (BES) experiments~~\cite{Nayak:2009wc,Aggarwal:2010wy,McDonald:2015tza} at the Relativistic Heavy Ion Collider (RHIC) as well as at the future Facility for Antiproton and Ion Research (FAIR). These experiments are expected to shed light on whether the phase transition from the baryon-rich QGP to the hadronic matter is a first-order one and the location of the critical end point in the QCD phase diagram if the phase transition is first-order.  To help understand what could happen in a baryon-rich QGP, we have recently used the Polyakov-Nambu-Jona-Lasinia (PNJL) model to study its spinodal instability~\cite{PhysRevC.93.035205}.  We have found via the linear response theory that the spinodal boundary in the temperature and density plane of the QCD phase diagram shrinks with increasing wave number of the unstable mode and is also reduced in the absence of the Polyakov loop.  In the small wave number or long wavelength limit, the spinodal boundary coincides with that determined from the isothermal spinodal instability in the thermodynamic approach.  We have further found that the quark vector interaction suppresses unstable modes of all wave numbers. For the wave number dependence of the growth rate of unstable modes, it initially increases with the wave number but decreases when the wave number is large. For the collisional effect from quark scattering, we have included it via the linearized Boltzmann equation and found it to decrease the growth rate of unstable modes of all wave numbers. In the present study, we continue the above work by investigating how unstable modes would grow if one goes beyond the linear response or small amplitude limit.  This is carried out by using the transport equation derived from the NJL model to study the time evolution of density fluctuations in a confined as well as in an expanding quark matter.  Specifically, we study the time evolution of higher-order density moments in the quark matter, the distribution of quark number in a sub-volume of the quark matter, the quark momentum anisotropy, and dilepton production rate from quark-antiiquark annihilation.  As shown below, these observables could serve as signatures for a first-order phase transition of the baryon-rich quark matter produced in heavy ion collisions. 

The paper is organized as follows: In the next section, we give a brief review on the NJL model and the transport equations based on its Lagrangian. The transport equations are solved by the test-particle method in Section III to study both the short and long time behavior of the spinodal instability of a quark matter in a periodic box. The same method is applied in Section IV to an expanding quark matter to study how density fluctuations are affected by the expansion of the system as in heavy ion collisions. Finally, a summary is given in Section V.  In the Appendix, we describe in detail the effect due to the finite grids used in the numerical calculations on the growth rate of unstable modes. 

\section{The NJL Lagragian And The Transport Model} 

The NJL Lagrangian containing only the scalar interaction for three quark flavors has the form~\cite{Bratovic:2012qs}:
\begin{eqnarray}
\label{LS}
&&\mathcal{L}^S_{NJL}=\bar{q}(i\not{\partial}-m_0)q+\frac{G_S}{2}\sum_{a=0}^{8}\bigg[(\bar{q}\lambda^aq)^2+(\bar{q}i\gamma_5\lambda^aq)^2\bigg]\nonumber\\
&&\quad-K\bigg[{\rm det}_f\bigg(\bar{q}(1+\gamma_5)q\bigg)+{\rm det}_f\bigg(\bar{q}(1-\gamma_5)q\bigg)\bigg],
\end{eqnarray}
where $q=(u,d,s)^T$, $m_0={\rm diag}(m_{0u}, m_{0d}, m_{0s})$ and $\lambda^a$ are the Gell-Mann matrices for $a=1,2 \cdots 8$, with $\lambda^0$ being the identity matrix multiplied by $\sqrt{2/3}$. The Lagrangian preserves $U(1)\times SU(N_f)_L\times SU(N_f)_R$ symmetry
but breaks the axial symmetry, which is broken in QCD by the axial anomaly, by the Kobayashi-Masakawa-t'Hooft (KMT) interaction given by the last term in  Eq. (\ref{LS})~\cite{'tHooft:1976fv}. The ${\rm det}_f$ in this term denotes the determinant in the flavor space~\cite{Buballa:2003qv}, that is 
\begin{eqnarray}
{\rm det}_f (\bar{q}\Gamma q)=\sum_{i,j,k}\varepsilon_{ijk}(\bar{u}\Gamma q_i)(\bar{d}\Gamma q_j)(\bar{s}\Gamma q_k),
\end{eqnarray}
where $\Gamma$ denotes either a Dirac gamma or the identity matrix. The determinantal term is responsible for obtaining the correct splitting in the masses of $\eta$ and $\eta^\prime$ mesons.

Because the NJL model is not renormalizable, a regularization scheme is required to remove infinities in the momentum integrations. In this study, we assume that all interactions are among quarks of 3-momenta with magnitudes below the cutoff momentum $\Lambda$.  Taking $\Lambda=0.6023~\mathrm{GeV}$, the values of the scalar coupling $G_S$ and the KMT interaction $K$ can be determined from fitting the pion mass, the kaon mass, and the pion decay constant, and their values are $G_S\Lambda^2=3.67$, and $K\Lambda^5=12.36$ if the current quark masses are taken to be $m_{0u}=m_{0d}=3.6$ and $m_{0s}=87$ MeV ~\cite{Holstein:1990ua}. 

A flavor-singlet vector interaction can be added to the NJL Lagrangian as follows:
\begin{equation}
\mathcal L^V_{NJL}=-G_V(\bar q\gamma^\mu q)^2,
\end{equation}
where the coupling strength $G_V$ is assumed to be independent of the temperature $T$ and the net quark chemical potential $\mu$.
The value of $G_V$ affects the order of quark matter phase transition. If $G_V$ is large, the first-order phase transition induced by the attractive scalar interaction could disappear~\cite{PhysRevC.93.035205}. In the present study, we treat it as a parameter to change the equation of state of quark matter. 

For describing the quark matter produced in a heavy ion collision, we use the Boltzmann (or transport) equations that can be derived from the NJL Lagrangian in terms of the non-equilibrium Green's functions for quarks and antiqaurks~\cite{Klevansky:1997wm}, and they are:
\begin{eqnarray}
\label{vlasov_q}
&&\partial_{X^0}f_a(X,\mathbf p)+\frac{p^{i\pm}}{E_{\mathbf p^\pm}}\partial_{X^i}f_a(X,\mathbf p)\nonumber\\
&&-\partial_{X^i}V_a^S(X)\frac{M_a}{E_{\mathbf p^\pm}}\partial_{p_i}f_a(X,\mathbf p)\mp\partial_{X^i}V^V_0(X)\partial_{p_i}f_a(X,\mathbf p)\nonumber\\
&&\mp\partial_{X^i}V^V_j(X)\frac{p^{j\pm}}{E_{\mathbf p^\pm}}\partial_{p_i}f_a(X,\mathbf p)=\mathcal C[f_a],
\end{eqnarray}
where $f_a(X,{\bf p})$ is the phase-space distribution function of quarks or antiquarks of flavor $a$. In the above, $\mathbf p^\pm\equiv\mathbf p\pm\mathbf V^V$ is the kinetic momentum with the subscript $+$ referring to quarks and $-$ referring to antiquarks, $V^V_\mu=-2G_V\sum_a\langle\bar q\gamma_\mu q\rangle_a$ is the vector potential, and $M_a=m_{0a}-V^S_a$ is the effective quark mass with $V^S_a=2G_S\langle\bar qq\rangle_a+2K\langle\bar qq\rangle_b\langle\bar qq\rangle_c$ being the scalar potential with $a\neq b\neq c$. 

The right hand side of Eq.(\ref{vlasov_q}),
\begin{widetext}
	\begin{eqnarray}
	\mathcal C[f_a]&\equiv& \sum_{bcd}\frac 1 {1+\delta_{ab}}\int\frac{d^3\mathbf p_b}{(2\pi)^32E_b}\frac{d^3\mathbf p_c}{(2\pi)^32E_c}\frac{d^3\mathbf p_d}{(2\pi)^32E_d}\frac{(2\pi)^4}{2E_a}\delta^4(p_a+p_b-p_c-p_d)\nonumber\\
	&&\times|\mathcal M_{ab}|^2\left[f_cf_d(1-f_a)(1-f_b)-f_af_b(1-f_c)(1-f_d)\right],
	\end{eqnarray}
\end{widetext}
is the collisional term that describe the scatterings among quarks and antiquarks, with the subscripts $a$, $b$, $c$, and $d$ now denoting not only the flavor but also the spin and color, and baryon charge (quark or anti-quark) of a parton. The above equation can be solved using the test particle method~\cite{Wong:1982prc} by expressing the distribution function in terms of the density of test particles, whose equations of motions are determined by the left hand side of Eq.(\ref{vlasov_q}), and they are
\begin{eqnarray}
\label{eom1}
\dot{\mathbf x}&=&\frac{\mathbf p^\pm}{E_{\mathbf{p}^\pm}},\\
\dot{\mathbf p}&=&\nabla V^S(\mathbf x)\frac{M}{E_{\mathbf{p}^\pm}}\pm\nabla V^V_0(\mathbf x)\pm\nabla V^V_j(\mathbf x)\frac{p^{j\pm}}{E_{\mathbf{p}^\pm}}.
\end{eqnarray}
The second equation in the above can also be written as
\begin{equation}
\label{eom3}
\dot{\mathbf p}^\pm=\nabla V^S(\mathbf x)\frac{M}{E_{\mathbf{p}}}\mp\dot\mathbf x\times\mathbf B\pm\mathbf E,
\end{equation}
where $\mathbf B=\nabla\times\mathbf V^V$ is the strong magnetic field and $\mathbf E=\partial_t\mathbf V^V+\nabla V^S$ is the strong electric field.

Besides the mean fields, test particles are also affected by collisions, which can be treated geometrically by generalizing the method of Ref.~\cite{Bertsch:1988ik} to use the particle scattering cross section $\sigma$ in the quark matter frame to check whether the impact parameter between two colliding particles is smaller than $\sqrt{\sigma/\pi}$ and if the two colliding particles pass through each other at the next time step during the evolution of the system.  For two particles of masses $m_A$ and $m_B$,  momenta ${\bf p}_A$ and ${\bf p}_B$, and energies $E_A$ and $E_B$, this cross section is related to the cross section in their center-of-mass frame $\sigma_{\rm CM}(\sqrt{s})$ with $s=(p_A+p_B)^2$ being the square of their invariant mass, which is the one used in Ref.~\cite{Bertsch:1988ik}, by
\begin{equation}
\sigma=\sigma_\mathrm{CM}(\sqrt{s})\frac{\sqrt{(s-(m_A+m_B)^2)(s-(m_A-m_B)^2)}}{2E_AE_B|\mathbf v_A-\mathbf v_B|}.
\end{equation}
In the above, ${\bf v}_A={\bf p}_A/E_A$ and ${\bf v}_B={\bf p}_B/E_B$ are the velocities of the two particles.  The 3-momenta of the two particles after the scattering are taken to be isotropic in their center-of-mass frame.  Because of the high quark baryon chemical potential considered in the present study, the Pauli blocking effect on scatterings is also included by checking the available phase space for the final states~\cite{Bertsch:1988ik}.  We have checked that the above treatment of parton scattering reproduces the expected scattering rate evaluated via direct numerical integrations.

\section{\uppercase {Quark matter in a box}}

This section serves as a bridge between the studies of the spinodal instabilities in the small and large amplitude limits. Although the case of small amplitude has already been discussed in Ref.~\cite{PhysRevC.93.035205}, we can develop an intuitive picture for how an initial sinusoidal fluctuation in a baryon-rich quark matter grows during the early stage of its time evolution from solving the Boltzmann equation in the test particle method as discussed in the previous section. For the large amplitude case, which also includes the growth of instabilities during the late stage, solving the Boltzamnn equation allows us to follow the whole phase separation process to see how dense clusters develop inside a box of initially uniform quark matter and finally lead to the formation of large scale structures.  It also provides the possibility to find the appropriate observables to characterize these structures.

\subsection{Small amplitude density fluctuations}

\begin{figure}[htbp]
	\vspace{-0cm}
	\includegraphics[width=0.45 \textwidth]{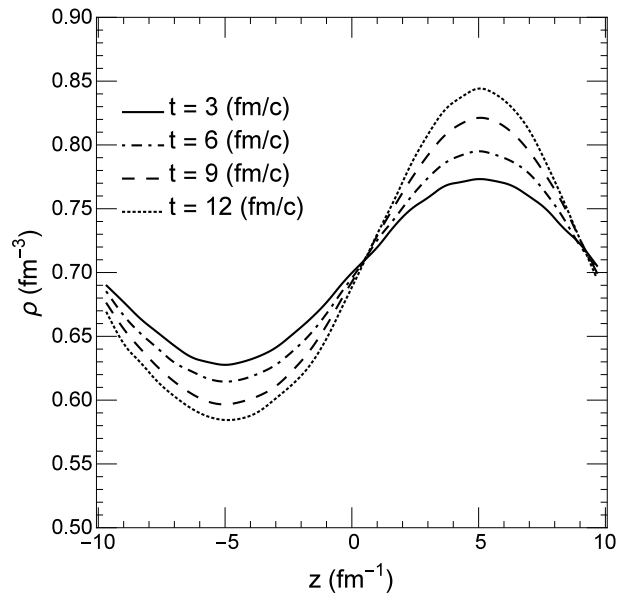}
	\caption{ Time evolution of an unstable density mode of wave number $k=0.31$ fm$^{-1}$.}
	\label{mode_grow}
\end{figure}

\begin{figure}[htbp]
	\vspace{-0cm}
	\includegraphics[width=0.45 \textwidth]{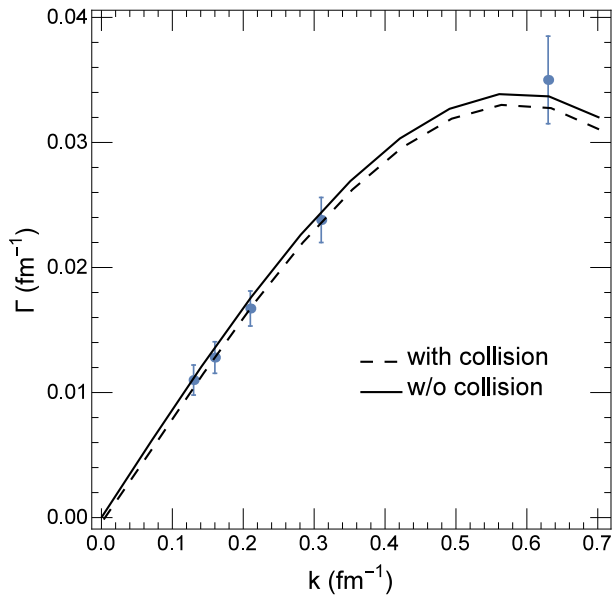}
	\caption{ Growth rates extracted from numerically solving the Boltzmann equation for unstable modes of wave numbers $k=0.63$, $0.31$, $0.21$, $0.16$, $0.13~\mathrm{fm}^{-1}$ for quark matter of density $\rho=0.7$ fm$^{-3}$ and temperature $T=45$ MeV. Analytical results from the linearized Boltzmann equation of Ref.~\cite{PhysRevC.93.035205} are shown by  solid and dashed curves for the cases with and without the collision term, respectively.}
	\label{box_check}
\end{figure}

We consider a quark matter that is confined in a cubic box with periodic boundary conditions.   The system is prepared by distributing many test particles inside the box according to the density of the system with their momenta given by the Fermi-Dirac distribution at certain temperature. We then study the growth of density fluctuations from an initial distribution with density and temperature corresponding to that inside the spinodal region.   Results obtained from solving the Boltzmann equation by following the classical motions of these test particles are compared with those obtained from the linear response theory in Ref.~\cite{PhysRevC.93.035205}. Specifically, we introduce an initial  density fluctuation that has a sinusoidal oscillation in the z direction, $\rho_\mathrm{ini}=\rho_0 (1+0.1\sin(2\pi z/L))$, where $\rho_0$ is the average initial density and $L$ is the length of the box with $L=10,~20,~30,~40,~50~\mathrm{fm}$ corresponding to  wave numbers $k=0.63,~0.31,~0.21,~0.16,~0.13~\mathrm{fm}^{-1}$, respectively. As an example, Fig.~\ref{mode_grow} shows how the amplitude of the sinusoidal wave grows with time in the case of $L=20~\mathrm{fm}$, the average density $\rho_0=0.7~\mathrm{fm}^{-3}$, and an initial temperature $T=45$ MeV. Since the amplitude of  density fluctuation at early times is expected to grow exponentially, it can be approximated by a hyperbolic cosine function of time, i.e.,
\begin{equation}
\delta\rho(t)=\delta\rho_0\mathrm{cosh}(\Gamma_k t),
\end{equation}
where $\Gamma_k$ is the growth rate and can be extracted directly from the numerical results, and they are shown in Fig. \ref{box_check} by solid circles. They are seen to agree very well with those obtained from an analytical calculation based on the linearized Boltzmann equation~\cite{PhysRevC.93.035205} after including the finite grid size effect as described in the Appendix, shown by the solid and dashed lines for the cases with and without the collision term in the Boltzmann equation, respectively. 
 
\subsection{Large amplitude density fluctuations}

\begin{figure}[h]
	\vspace{-0cm}
	\centering
	\includegraphics[width=0.45 \textwidth]{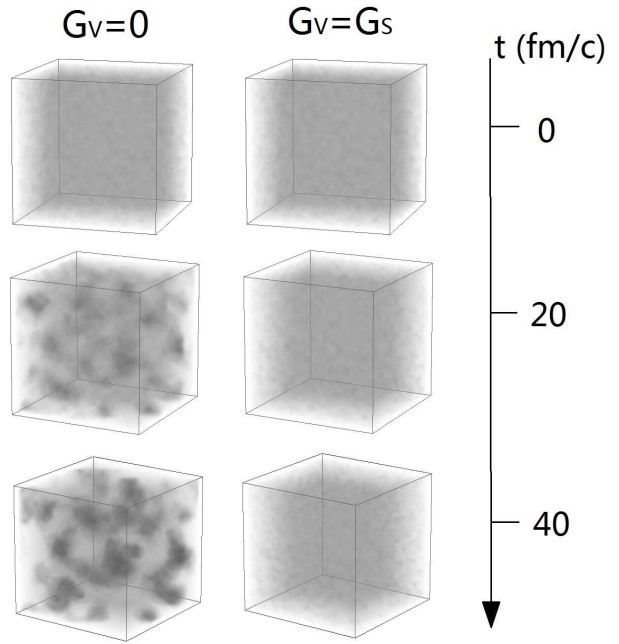}
	\caption{Time evolution of density distribution in a quark matter of temperature $T=20$ MeV and net quark density $n_q=0.5~\mathrm{fm}^{-3}$ for the cases of $G_V=0$ (left column) and $G_V=G_S$ (right column).}
	\label{3d_cube_cll_n_05_T_020}
\end{figure}

\begin{figure}[h]
	\vspace{-0cm}
	\centering
	\includegraphics[width=0.45 \textwidth]{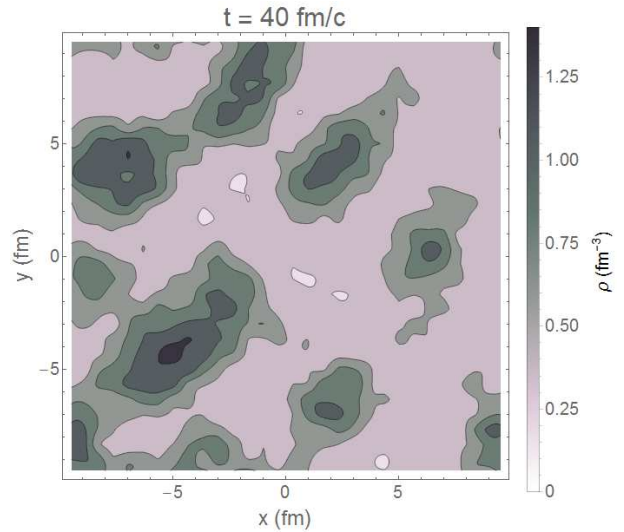}
	\caption{Cross sectional view of  density distribution on the $z=0$ plane at $t=40~\mathrm{fm}/c$ for the case $G_V=0$ with a first-order phase transition.}
	\label{rho40}
\end{figure}

To study how density fluctuations emerge and grow, we compare results from two calculations based on the  same initial conditions but with and without the spinodal instability in the equation of state. This is achieved by introducing a vector interaction in the NJL model, which is known to move the state of a quark matter from inside the spinodal region to the outside if its strength is sufficiently large~\cite{PhysRevC.93.035205}. For example, for a quark matter of temperature $T_0=$ 20 MeV and net quark density $\rho_0=0.5~\mathrm{fm}^{-3}$,  the spinodal region disappears if the vector coupling $G_V$ has the same value as the scalar coupling $G_S$, although the state of the quark matter is well inside the spinodal instability region for $G_V=0$.  

Figure~\ref{3d_cube_cll_n_05_T_020} shows the time evolution of the density distribution in a box of size $20\times20\times20~\mathrm{fm}^3$ for the two cases of $G_V=0$ (left column) and $G_V=G_S$ (right column), with the darker color denoting the high density regions and the lighter color denoting the low density regions.  Although the system is initially uniform in space, some dense spots are present due to statistical fluctuations as a result of finite number of test particles used in the calculation.  In the case of $G_V=G_S$ without a first-order phase transition or spinodal instability, the density distribution in the box remains unchanged with time as shown in the right column.  This changes dramatically, however, for the case of $G_V=0$. Due to the spinodal instability, the initial dense spots act like "seeds", which create several small low pressure regions that attract nearby partons and lead to the formation of many clusters at $t=20$ fm$/c$.  These  clusters further grow in size by connecting with each other and form stable large structures at $t=40$ fm$/c$, when the system clearly separates into two phases of matter with one of high density and the other of low density.  

\begin{figure}[h]
	\vspace{-0cm}
	\centering
	\includegraphics[width=0.45 \textwidth]{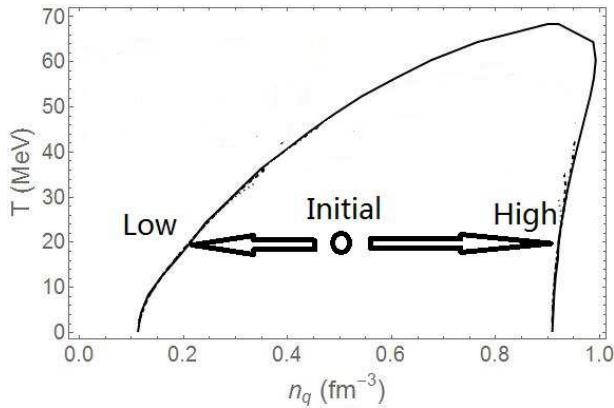}
	\caption{Demonstration of the phase separation in the phase diagram.}
	\label{demon}
\end{figure}

A clearer picture can be obtained by taking a cross sectional view on the $z=0$ plane as shown by the density distribution contours in Fig. \ref{rho40}. The two phases are now distinguishable with the dilute phase having a density of about $0.25~\mathrm{fm}^{-3}$ and  the dense phase having a density of about $1.0~\mathrm{fm}^{-3}$. According to the phase diagram in Fig. \ref{demon}, the initial location of the system is indicated by the circle inside the spinodal region. During the phase separation, the location of most part of the system moves towards the left boundary of the spinodal instability region that has a density of about $0.2~\mathrm{fm}^{-3}$, while that of the small part of the system moves towards the right boundary of the spinodal instability region that has a density of about $0.9~\mathrm{fm}^{-3}$, consistent with the picture shown by the density evolution.

\begin{figure}[h]
	\vspace{-0cm}
	\centering
	\includegraphics[width=0.45 \textwidth]{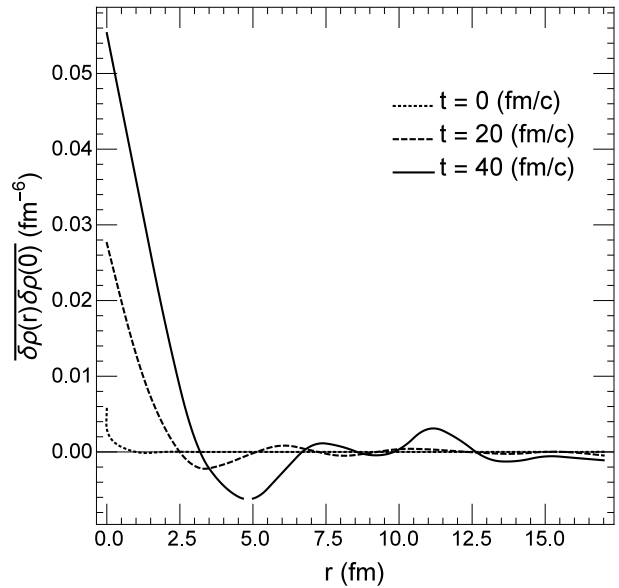}
	\caption{ Time evolution of the density-density correlation function in a quark matter of temperature $T=20$ MeV and average net quark density $n_q=0.5~\mathrm{fm}^{-3}$ inside the spinodal region.}
	\label{cor_T_020_n_05}
\end{figure}

As the large scale structure forms, we expect the density-density correlation $\overline{\rho(r)\rho(0)}$ to get stronger and the correlation length to become larger. This is indeed the case as shown in Fig.~\ref{cor_T_020_n_05}, where it is seen that both the amplitude of the correlation function and the correlation length increases with time. 

\begin{figure}[h]
	\vspace{-0cm}
	\centering
	\includegraphics[width=0.45 \textwidth]{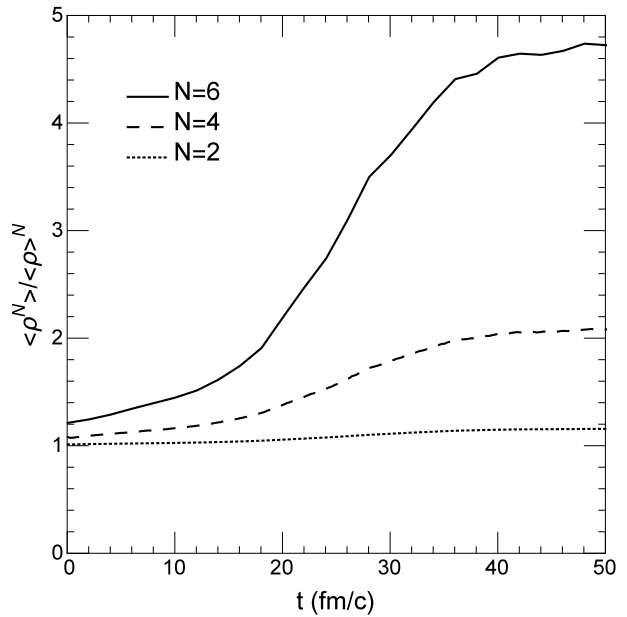}
	\caption{ Time evolution of the scaled density moments in a quark matter of temperature $T=20$ MeV and average net quark density $n_q=0.5~\mathrm{fm}^{-3}$ inside the spinodal region.}
	\label{moment_box_T_020_n_05}
\end{figure}

The density fluctuations can also be quantified by the scaled density moments $\langle \rho^N\rangle/\langle \rho\rangle^N$~\cite{PhysRevC.87.054903}, where
\begin{equation}
\langle\rho^N\rangle\equiv\frac{\int d^3\mathbf r \rho(\mathbf r)^{N+1}}{\int d^3\mathbf r \rho(\mathbf r)}.
\end{equation}
This quantity is scale invariant since its value remains unchanged under a scale transfomation $\mathbf r\to\lambda\mathbf r$, where $\lambda$ can be any positive number. The scaled density moments are all equal to one for a uniform density distribution but become greater than one as the density fluctuations grow. In Fig.~\ref{moment_box_T_020_n_05}, we show by dotted, dashed, and solid lines the scaled density moments for $N=2$, 4 and 6, respectively. Our results show that the scaled moments increase during the phase separation and reach their saturated values at about $t=40$ fm$/c$, when the phase separation almost ends. Also, moments with larger $N$ increase faster and saturate at larger values. The final saturation values can be  estimated as follows.  For a system of an initial density $\rho_0$ that  separates into two phases of density $\rho_1$ and $\rho_2$ with volumes $V_1$ and $V_2$, respectively, the scaled density moments are then
\begin{equation}
\frac{\langle\rho^N\rangle}{\langle\rho\rangle^N}=\frac{\rho_1^{N+1}V_1+\rho_2^{N+1}V_2}{\left(\rho_1^2 V_1+\rho_2^2 V_2\right)^N/\left(\rho_1V_1+\rho_2V_2\right)^{N-1}}.
\end{equation}
Using the condition of particle number conservation
\begin{equation}
\rho_1V_1+\rho_2V_2=\rho_0(V_1+V_2),
\end{equation}
the scaled density moments  after the phase separation is thus
\begin{equation}
\frac{\langle\rho^N\rangle}{\langle\rho\rangle^N}=\frac{[\rho_1^{N+1}(\rho_2-\rho_0)+\rho_2^{N+1}(\rho_0-\rho_1)][\rho_0(\rho_2-\rho_1)]^{N-1}}{[\rho_1^{2}(\rho_2-\rho_0)+\rho_2^{2}(\rho_0-\rho_1)]^N}.
\end{equation}
For our case of $\rho_0=0.5~\mathrm{fm}^{-3}$, $\rho_1\approx0.25~\mathrm{fm}^{-3}$, and $\rho_2\approx1.0~\mathrm{fm}^{-3}$, we have  $\langle\rho^2\rangle/\langle\rho\rangle^2\approx 1.22$,$\langle\rho^4\rangle/\langle\rho\rangle^4\approx 2.11$, and $\langle\rho^6\rangle/\langle\rho\rangle^6\approx 3.75$, which are close to the final saturation values shown in Fig.~\ref{moment_box_T_020_n_05}.

Other quantities of interest are the skewness and kurtosis of the particle multiplicity distribution, which were proposed as possible signals for the critical phenomena~\cite{Stephanov:2009prl} and have been studied in the beam energy scan experiments at RHIC\cite{Nayak:2009wc,Aggarwal:2010wy}. They are defined as follows:
\begin{eqnarray}
\mathrm{skewness}&\equiv&\frac{\langle\delta N_q^3\rangle}{\langle\delta N_q^2\rangle^{3/2}}, \nonumber\\
\mathrm{kurtosis}&\equiv&\frac{\langle\delta N_q^4\rangle}{\langle\delta N_q^2\rangle^2}-3.
\end{eqnarray}
Both quantities characterize how far an event-by-event multiplicity distribution deviates from a normal distribution. A positive skewness means a long tail on the right side of the distribution, i.e., most events have the net quark number below the mean value, while some events have an extreme high net quark number. A positive kurtosis implies a sharper peak than the peak in a normal distribution, while a negative kurtosis corresponds to a flatter one. Theoretical calculations based on the grand canonical picture predict that both quantities diverge with the correlation length when a system approaches its critical point~\cite{Stephanov:2009prl}, with the kurtosis diverging faster than the skewness. Therefore, they have thus been suggested as the signals for the existence of a critical end point in the QCD phase diagram.

\begin{figure}[h]
	\vspace{-0cm}
	\includegraphics[width=0.45 \textwidth]{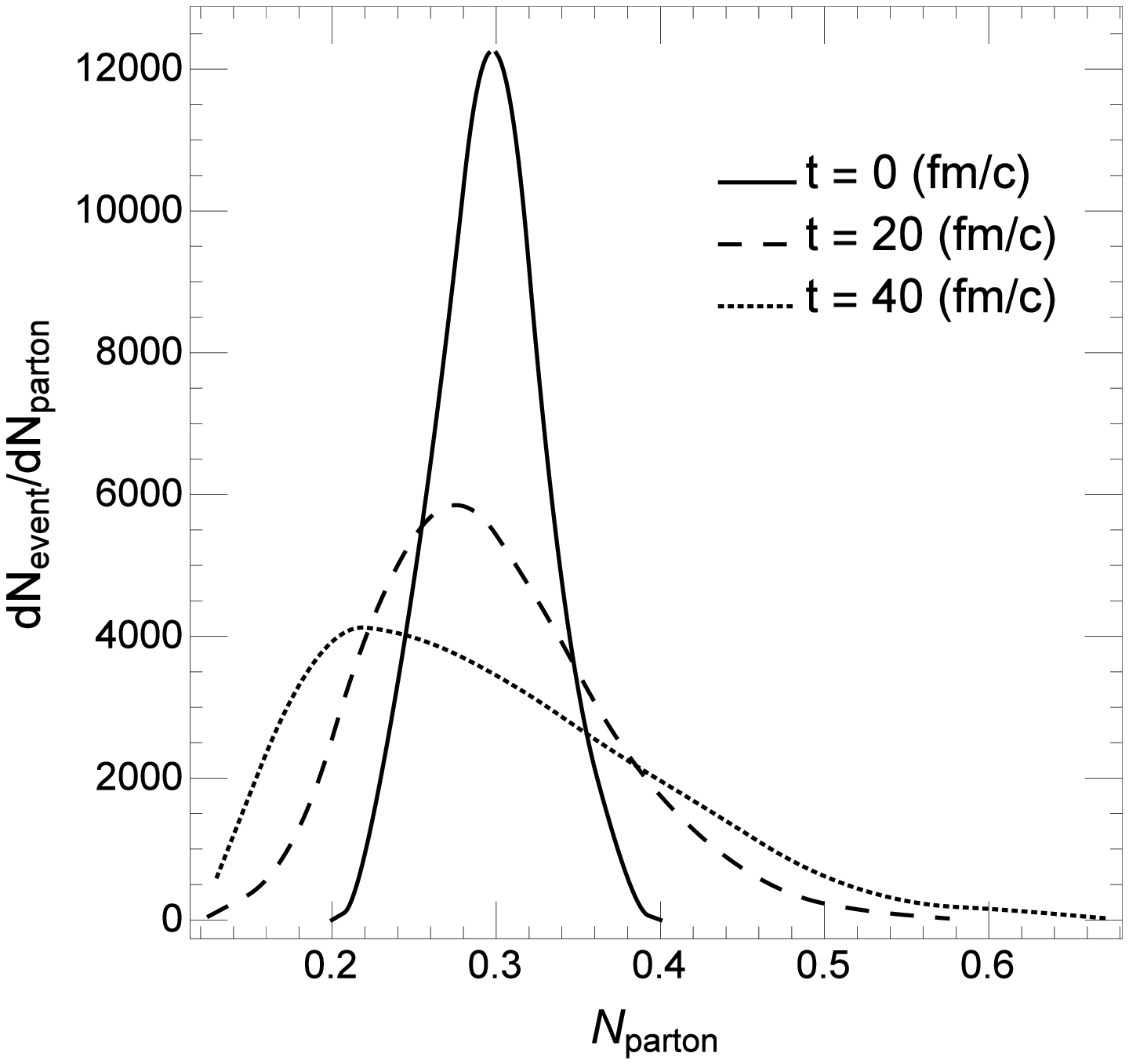}
	\includegraphics[width=0.45 \textwidth]{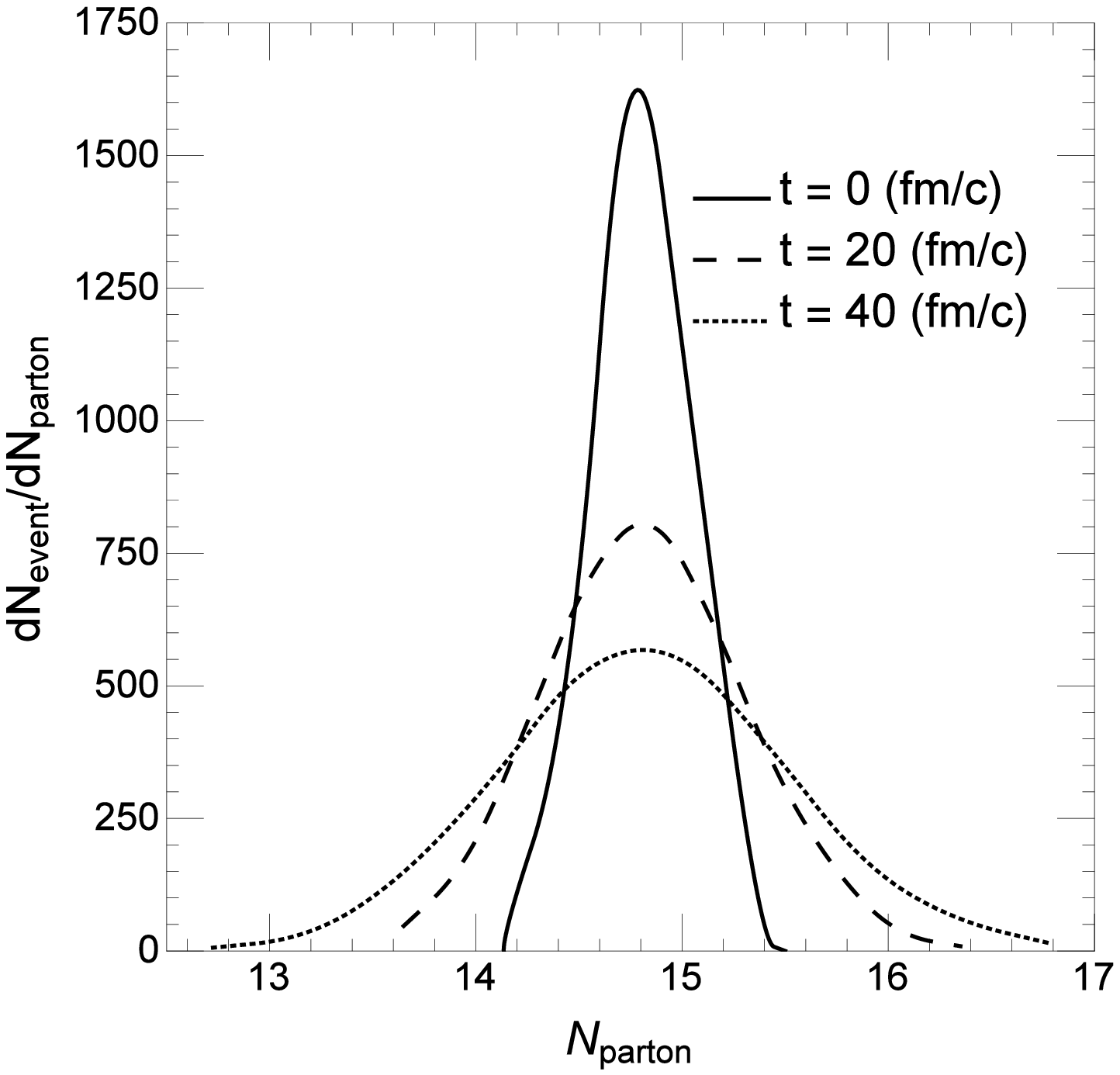}
	\caption{ Time evolution of the event-by-event distribution of the number of quarks  in a sub-volume of size 0.6 fm$^3$ (upper window) and 30 fm$^3$ (lower window) for a quark matter of temperature $T=20$ MeV and average net quark density $n_q=0.5~\mathrm{fm}^{-3}$ inside the spinodal region. The total number of events is 1000.}
	\label{box_event_by_event}
\end{figure}

To be consistent with the grand canonical picture, we consider quarks in a sub-volume of the box in our study, such as its central cell,  and treat the remaining part as the reservoir.  When the system is initially inside the spinodal instability region, quarks in the reservoir can sometimes move into the sub-volume, but in most of the times quarks would leave from the sub-volume to the reservoir.  The number of quarks inside this sub-volume thus varies drastically from event to event, leading to large values for the skewness and kurtosis in its event-by-event distribution. In Figs.~\ref{box_event_by_event}, we show the event-by-event distribution of the number of quarks in the central cell  from 1000 events at $t=0$, $20$, and $40$ fm$/c$  by the solid, dashed and dotted lines, respectively, for the two cases of sub-volume of size $0.6~\mathrm{fm}^3$ (upper window) and $30~\mathrm{fm}^3$ (lower window).  The upper window of Fig.~\ref{box_event_by_event} clearly shows that the distribution  for the small sub-volume becomes asymmetric as time increases, starting with an initial skewness of 0.11 and increasing to 0.60 at 20 fm$/c$ and 0.75 at 40 fm/$c$. This feature is absent in the lower window of Fig.~\ref{box_event_by_event} for the larger sub-volume, where the distribution remains essentially symmetric with increasing time, with the skewness changing slowly  from -0.001 (t=0) to 0.086 (t=20 fm$/c$) and 0.132 (t=40 fm$/c$), and there is no apparent increase or decrease in the kurtosis.

\section{\uppercase{expanding quark matter}}

\subsection{Blast wave initial conditions}

To study how large density fluctuations due to the spinodal instability as a result of a first-order phase transition obtained from the box calculation in the previous section are affected by the expansion of the system as in a heavy ion collision, we carry out a dynamical calculation using the transport model that includes parton scatterings besides the mean-field potentials described in Section II. For the initial parton distributions, their positions are taken to follow that of a spherical Wood-Saxon form:
\begin{equation}
\rho(r)=\frac{\rho_0}{1+\exp((r-R)/a)}
\end{equation}
with a radius $R=5~\mathrm{fm}$ and a surface thickness parameter $a=0.5~\mathrm{fm}$, similar to that expected from a central Au+Au collisions. The momenta of these patons are again taken to be that of a Fermi-Dirac distribution at certain temperature. Calculations are then carried out with  two different equations of state with and without a  first-order phase transition, which can be realized by adjusting the coupling strength for the vector interaction.

\begin{figure}[h]
	\vspace{-0cm}
	\centering
	\includegraphics[width=0.45 \textwidth]{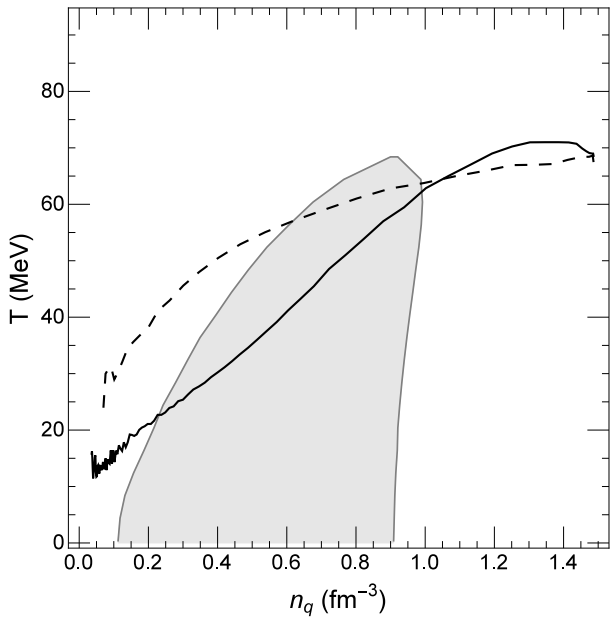}
	\caption{Phase trajectory of the central cell of an expanding quark matter for the two cases with (solid line) and without (dashed line) a first-order phase transition using the blast wave initial conditions. The spinodal region is shown by the gray color.}
	\label{trajectory}
\end{figure}

\begin{figure}[h]
	\vspace{-0cm}
	\centering
	\includegraphics[width=0.45 \textwidth]{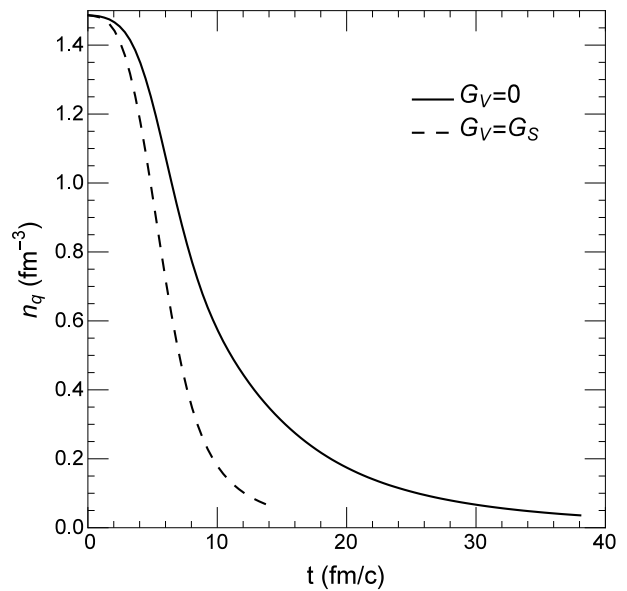}
	\caption{ Time evolution of the density of the central cell of  an expanding quark matter for the two cases with (solid line) and without (dashed line) a first-order phase transition.}
	\label{trajectory_t}
\end{figure}

To see how the expanding system goes into the spinodal region in the QCD phase diagram, we first study the time evolution of the temperature and net quark density in the central volume of 42.875 fm$^3$, which has an initial density $\rho_0=1.5~\mathrm{fm}^3$ and temperature $T=70$ MeV, and trace its phase trajectory as shown in Fig.~\ref{trajectory} for the two cases with (solid line) and without (dashed line) a phase transition. Although the quark matter described by the transport model may not always be in perfect thermal equilibrium, we approximate its  temperature  by that of an equilibrated one that has the same energy density and  net quark density  in the NJL model.  As expected, the quark matter with a first-order phase transition (solid curve) enters the spinodal instability region, which is shown by the gray color, at about $6.5~\mathrm{fm}/c$ and leaves the region at about $17.4~\mathrm{fm}/c$ after spending about $10~\mathrm{fm}/c$ inside this region. How the central density decreases with time is shown by the solid line in Fig.~\ref{trajectory_t}, which is seen to decrease slower than in the case without a first-order phase transition shown by the dashed line obtained with $G_V=G_S$

\begin{figure}[h]
	\vspace{-0cm}
	\includegraphics[width=0.45 \textwidth]{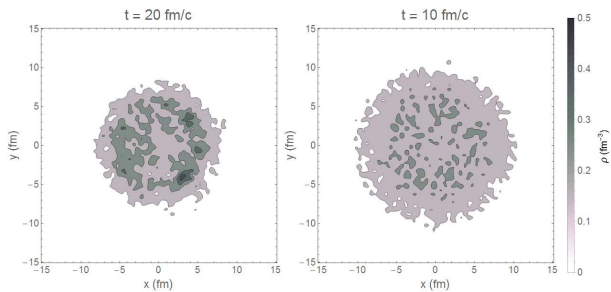}
	\caption{Density distributions of an expanding quark matter on the $z=0$ plane at $t=20$ fm/$c$ for the case with a first-order phase transition (left window) and at $t=10$ fm/$c$ for the case without a first-order phase transition (right window).}
	\label{Gv_0_T_070_n_15_t_20}
\end{figure}

\begin{figure}[h]
	\vspace{-0cm}
	\centering
	\includegraphics[width=0.45 \textwidth]{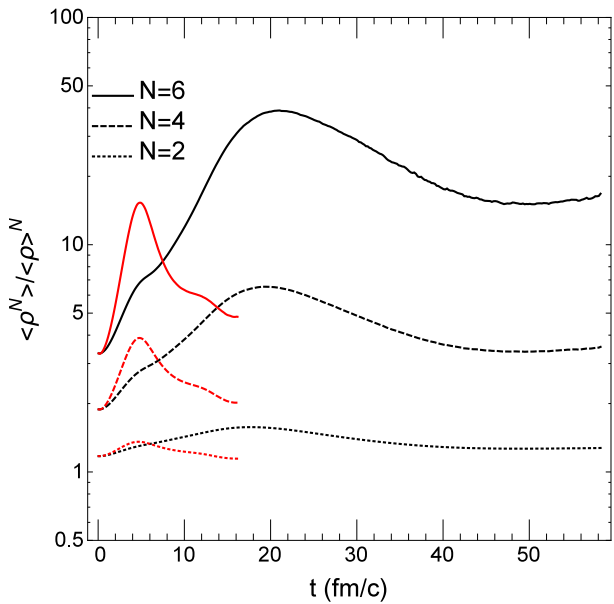}
	\caption{(Color Online). Scaled density moments  as  functions of time  for the cases with (black lines) and without (red lines) a first-order phase transition.}
	\label{moment}
\end{figure}

The density fluctuations can  be seen from the density distribution on a plane such as the one at $z=0$ shown in Fig.~\ref{Gv_0_T_070_n_15_t_20}. The left window shows the density distribution at $t=20$ fm/$c$ for the case with a first-order phase transition, while the right window shows that at $t=10$ fm/$c$ for the case without a first-order phase transition, when the density of the central cell is about $0.2$ fm$^{-3}$ in both cases. Although density clumps appear in both cases, those in the one with a first-order phase transition are significantly larger. As in the case of quark matter in a box, we can quantify the density fluctuations by the scaled density moments~\cite{PhysRevC.82.034902}.   They are shown  in Fig.~\ref{moment} by the black and red  lines for the cases with and without a first-order phase transition, respectively. The dotted, dashed, and  solid lines are for $N=2$, $4$, and $6$, respectively. In both cases, the scaled density moments  first increase and then decrease with time. In the case without a first-order phase transition, this is caused by the fast increase of the surface of the quark matter and the quick deviation from its initial smooth Wood-Saxon density distribution.  To the contrary, the scaled density moments in the case with a first-order phase transition becomes much larger with time and only decreases slightly afterwards, reflecting the effect due to density clumps that distribute randomly inside the expanding  quark matter. Therefore, the saturated  scaled density moments, which are larger for larger $N$, can be regarded as  signals for a first-order phase transition in a baryon-rich quark matter~\cite{PhysRevC.87.054903}.

\begin{figure}[htbp]
	\vspace{-0cm}
	\includegraphics[width=0.45 \textwidth]{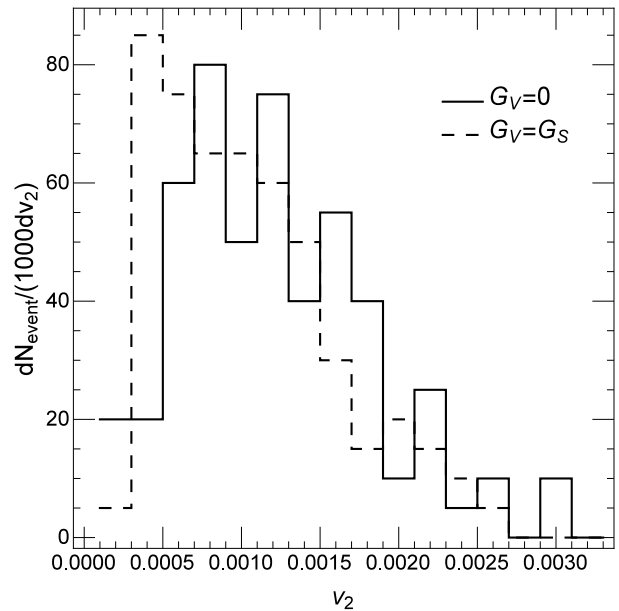}		
	\includegraphics[width=0.45 \textwidth]{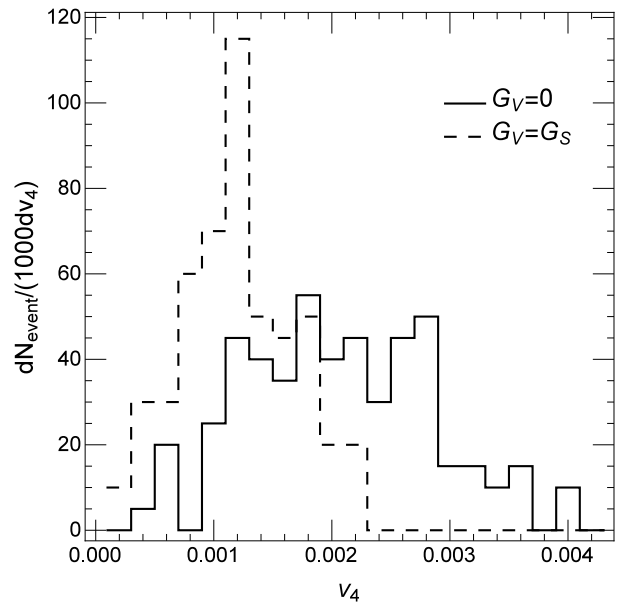}
	\caption{ Final  anisotropic flow coefficients $v_2$ (upper window) and $v_4$ (lower window) distributions for 100 events of an expanding quark matter with the same blast wave initial conditions.}
	\label{v2fin}
\end{figure}

Since density fluctuations can lead to spatial anisotropy even in central heavy ion collisions, it has been suggested that they may  affect the anisotropic flows in the transverse plane~\cite{Herold:2014uva,PhysRevC.89.034901}.  The latter are defined by the coefficients $v_n$ in the expansion of the transverse momentum distribution $f(p_T,\phi)$ as a Fourier series in the azimuthal angle $\phi$,
\begin{eqnarray}
f(p_T,\phi)=\frac{N(p_T)}{2\pi}\{1+2\sum_{n=1}^\infty v_n(p_T)\cos[n(\phi-\psi_n)]\},
\end{eqnarray}
where $\psi_n$ is the event plane angle~\cite{PhysRevC.82.034913}.   To calculate the anisotropic flow coefficients, we use the two particle cumulant method~\cite{PhysRevC.44.1091,PhysRevC.64.054901}, namely, $v_n\{2\}=\sqrt{\langle\cos(n\Delta\phi)\rangle}$ by averaging over all  particle pairs in an event.  We have calculated $v_2\{2\}$ and $v_4\{2\}$ for 100 events of an expanding quark matter with the same blast wave initial conditions, and their final event distributions  are shown, respectively, in the upper and lower windows of Fig. \ref{v2fin} with the solid and dashed lines  for the cases with and without first order phase transition, respectively. Both distributions peak at a larger value for the case with a first-order phase transition, particularly for $v_4$, thus providing a plausible signal for the first-oder phase transition.  However, the values of the fluctuation induced $v_2$ and $v_4$ are much smaller than those in non-central heavy ion collisions.

\begin{figure}[h]
	\hspace{2cm}
	\includegraphics[width=0.45\textwidth]{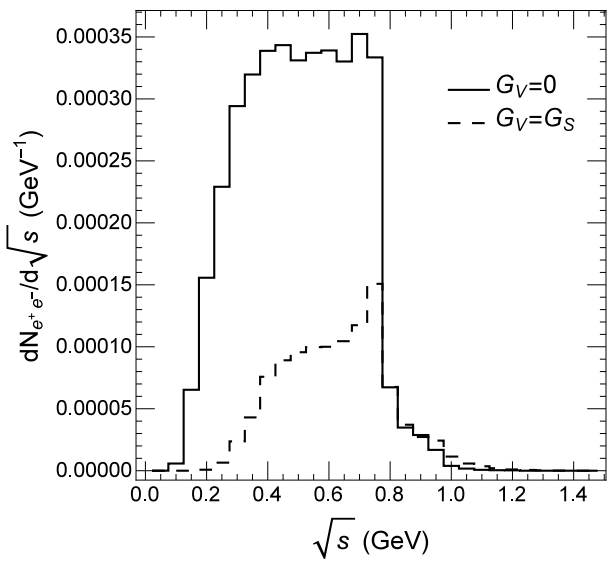}
	\caption{ Dilepton yield as a function of the invariant mass $\sqrt s$ for the cases with (solid line) and without (dashed line) a first-order phase transition in an expanding quark matter with the blast wave initial conditions.}
	\label{electron}
\end{figure}

We have also studied the effect of density fluctuations on dilepton production from a quark matter. Since the dilepton production rate is proportional to the square of parton density, more dileptons are produced when the density fluctuation is large. Also, a longer partonic phase as a result of a first-order phase transition would increase the depletion yield as well. As usually done in studying dilepton production in heavy ion collisions~\cite{Xiong:1990bg}, we use the perturbative approach to calculate the dilepton yield from the quark-antiquark scattering by neglecting its effect on the dynamics of the expanding quark matter.  Using the dilepton production cross section,
\begin{eqnarray}
\sigma_{q\bar q\to e^+e^-}&=&\frac{4\pi\alpha^2}{3s}\sqrt{\frac{1-4m_e^2/s}{1-4m_q^2/s}}\nonumber\\
&&\times\bigg(1+2\frac{m_e^2+m_q^2}{s}+4\frac{m_e^2m_q^2}{s^2}\bigg),
\end{eqnarray}
where $s=(p_{e^-}+p_{e^+})^2$ is the square of the dilepton invariant mass, we have calculated the dilepton invariant mass spectrum from the expanding  quark matter, and they are shown in Fig.~\ref{electron} by the solid and dashed lines for the cases with and without first-order phase transition, respectively.  As expected, more dileptions are produced from the  quark matter with a first-order phase transition. We note the dilepton invariant mass spectrum peaks at $\sqrt s\approx 0.5$ GeV with the peak value being about $3.5\times 10^{-4}$ GeV$^{-1}$, which is comparable with the result obtained from a hadronic transport model \cite{Galatyuk:2015pkq}.   This enhancement in dilepton production may thus be detectable in experiments.   We also note that most dileptons are produced from  quark-antiquark annihilation as very few pions are present in the system due to the low phase transition temperature $T_c$ in the SU(3) NJL model.  

\subsection{AMPT initial conditions}

In this subsection, we use a more realistic initial parton distribution for heavy ion collisions. Specifically, the initial partons are  obtained from a multiphase transport (AMPT) model with string melting~\cite{Lin:2004en}  that uses the heavy ion jet interaction generator (HIJING)~\cite{PhysRevD.43.104,PhysRevD.44.3501,Gyulassy1994307} as the input. This model includes not only the mini-jet partons from initial hard collisions but also hadrons produced from excited strings, which are projectile and target nucleons that have suffered interactions, by converting them to partons according to the flavor and spin structures of their valence quarks. In particular, a meson is converted to a quark and an anti-quark, while a baryon is first converted to a quark and a diquark, and the diquark is then decomposed into two quarks. The quark masses are taken to be $m_u=5.6$, $m_d=9.9$, and $m_s=199~\mathrm{MeV}/c^2$ as in the PYTHIA program \cite{Sjöstrand199474}. The above two-body decomposition is isotropic in the rest frame of the parent hadron or diquark.  These partons are produced after a formation time of $t_f=E_H/m^2_{T,H}$, with $E_H$ and $m_{T,H}$ denoting, respectively, the energy and transverse mass of the parent hadron.  We obtain these partons  as the initial conditions for our study of an expanding quark matter by running the AMPT program with vanishing parton scattering cross sections in Zhang's parton cascade (ZPC)\cite{Zhang1998193} and with the hadronic afterburner based on a relativistic transport (ART) \cite{PhysRevC.52.2037, ART2001} turned off. Using the partons from Au+Au collisions at zero impact parameter and a center-of-mass energy $\sqrt{s_{NN}}=2.5$ GeV as the initial distribution, we have found that some parts of the system go through the spinodal region when the SU(3) NJL model with $G_V=0$ is used in the Boltzmann equation and in constructing the phase diagram. 

\begin{figure}[h]
	\hspace{2cm}
	\includegraphics[width=0.45 \textwidth]{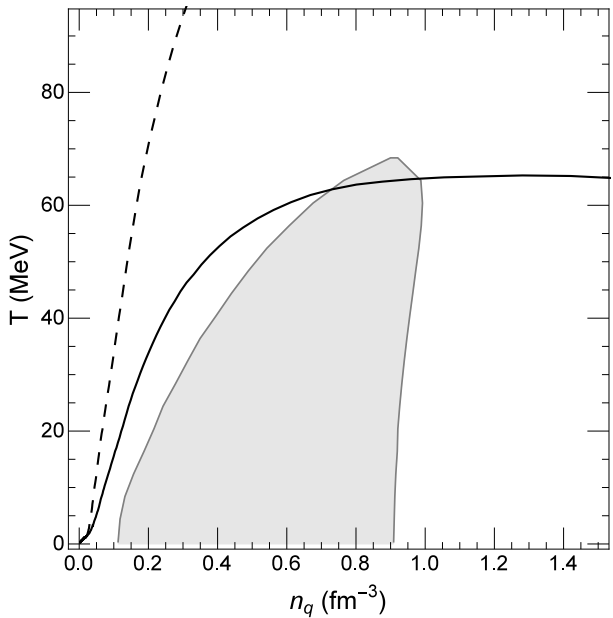}
	\caption{ Phase trajectories of the central part of an expanding quark matter for the cases with (solid line) and without (dashed line) a first-order phase transition using the initial parton distribution from the AMPT model. The spinodal region is shown by the gray color.}
	\label{trajectory_zpc}
\end{figure}

As shown by the solid line in Fig.~\ref{trajectory_zpc}, the trajectory of the central part of the system goes into the spinodal instability region at about $4.4~\mathrm{fm}/c$ after expansion, and moves out of  this region at about $5~\mathrm{fm}/c$. Although $0.6~\mathrm{fm}/c$ is too short for the spinodal instability to develop in the central part of the  quark matter, its other parts  may stay longer in the spinodal instability region  due to both the spatial distribution of initial partons and the correlations between their rapidities and longitudinal ($z$) coordinates.

\begin{figure}[h]
	\vspace{-0cm}
	\centering
	\includegraphics[width=0.45 \textwidth]{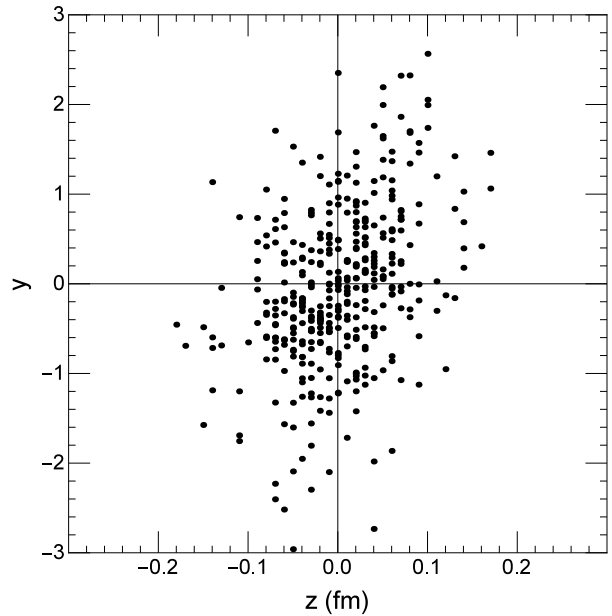}
	\caption{ Rapidity and longitudinal coordinate correlations of initial partons from the AMPT model for central Au+Au collisions at $\sqrt{s_{NN}}=2.5~\mathrm{GeV}$.}
	\label{ini_corr_zpc_25}
\end{figure}

\begin{figure}[h]
	\vspace{-0cm}
	\centering
	\includegraphics[width=0.45 \textwidth]{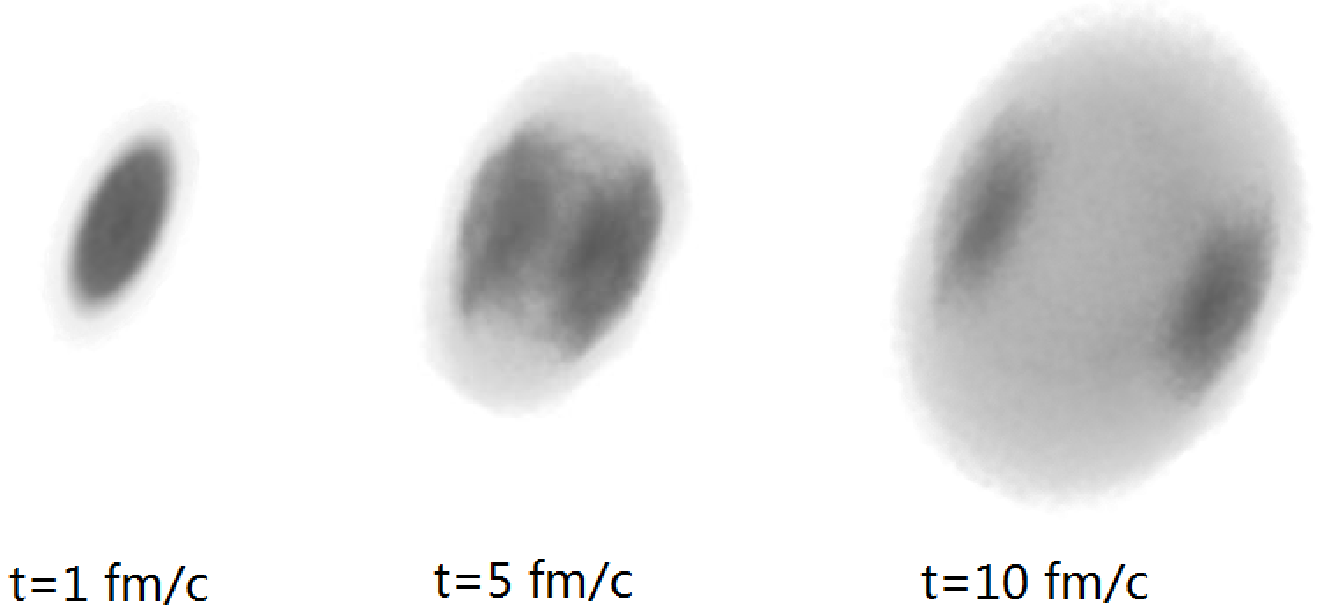}
	\includegraphics[width=0.45 \textwidth]{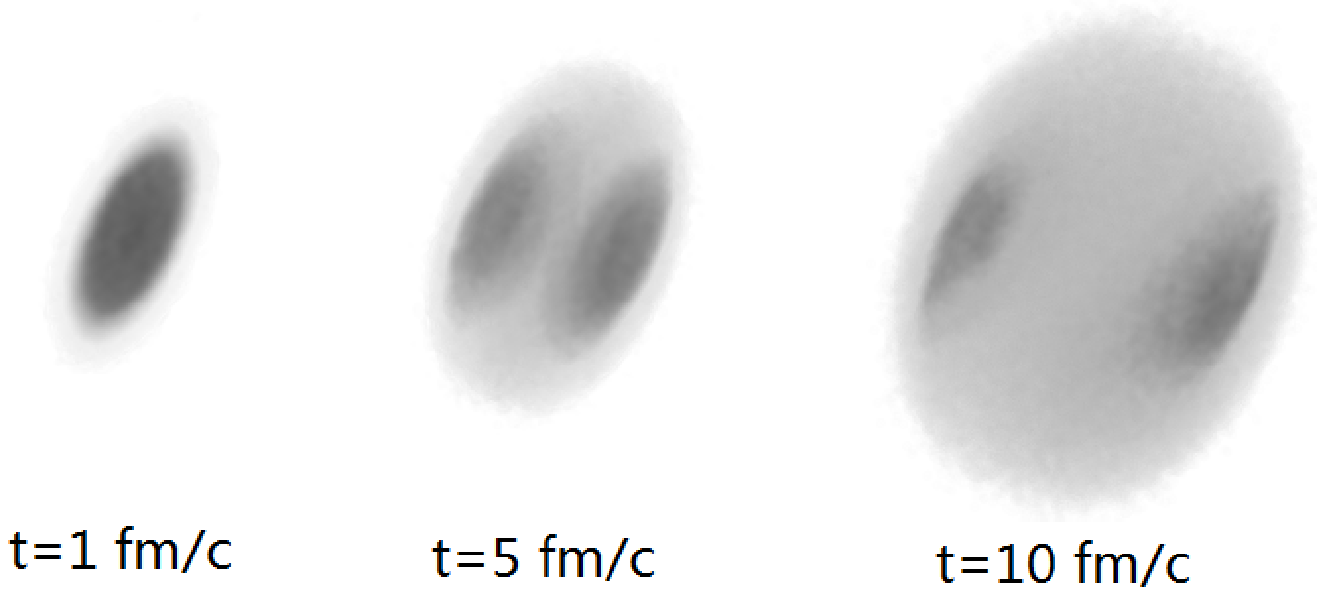}
	\includegraphics[width=0.45 \textwidth]{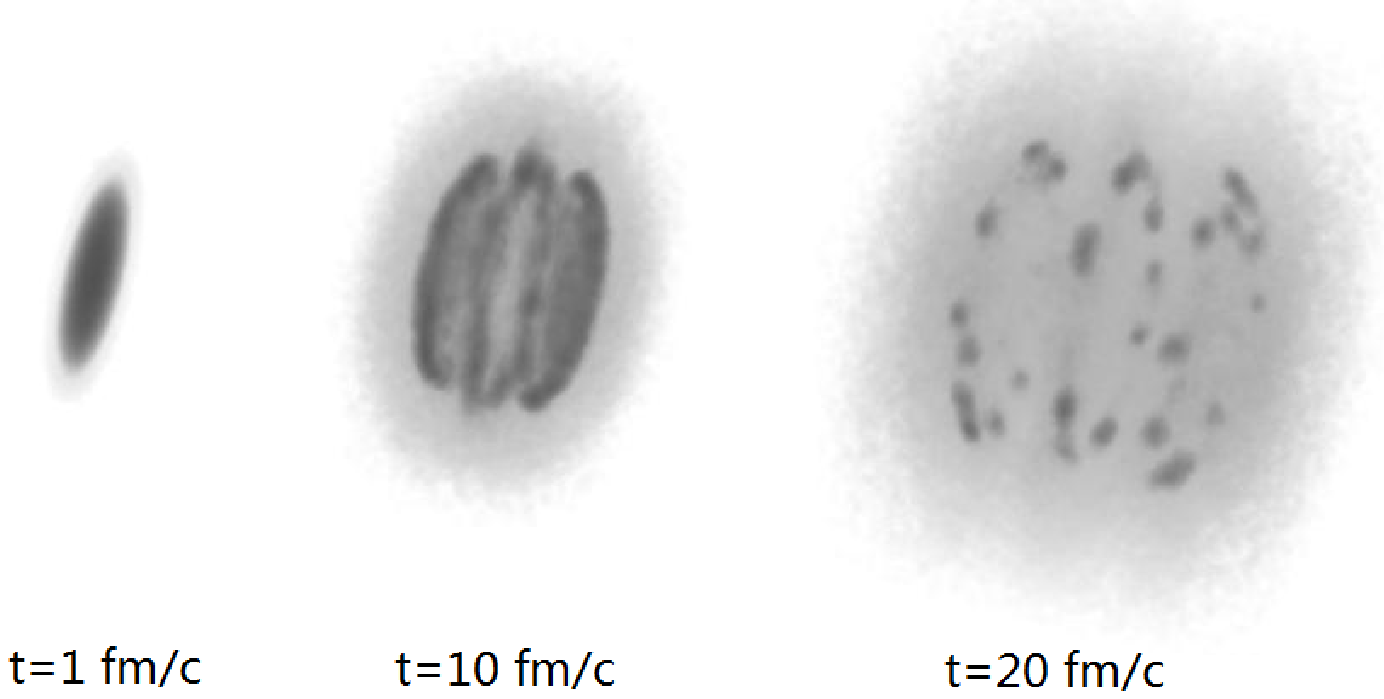}
	\caption{ Time evolution of the density distributions in central Au+Au collisions at $\sqrt{s_{NN}}=2.5~\mathrm{GeV}$ using initial conditions from the AMPT for the cases of free streaming (upper row) and including quark scattering as well as mean fields from the NJL model with $G_V=G_S$ (middle row) and $G_V=0$ (lower row).}
	\label{3d_zpc_025_free}
\end{figure}

\begin{figure}[h]
	\vspace{-0cm}
	 \includegraphics[width=0.45 \textwidth]{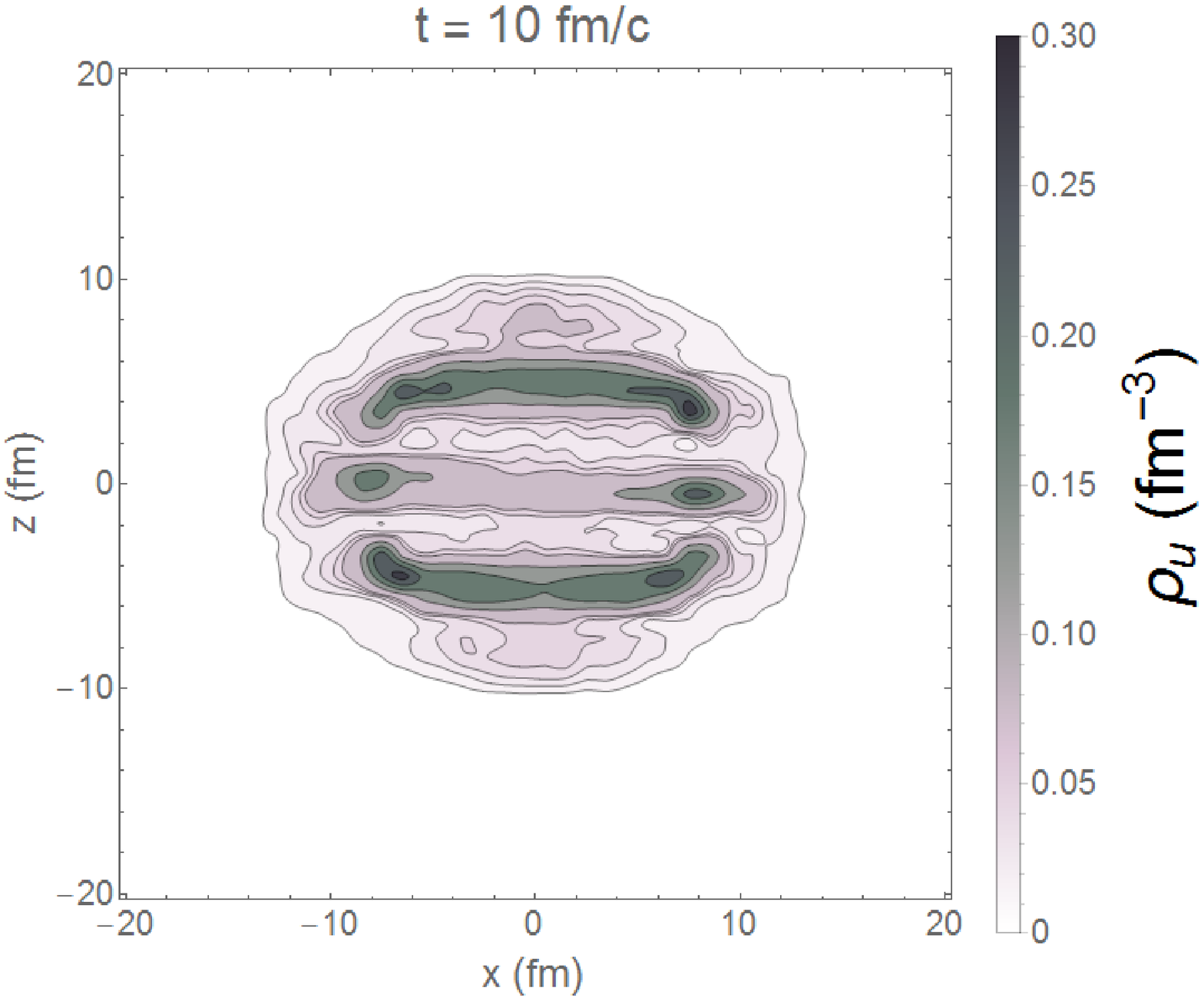}
		\includegraphics[width=0.45 \textwidth]{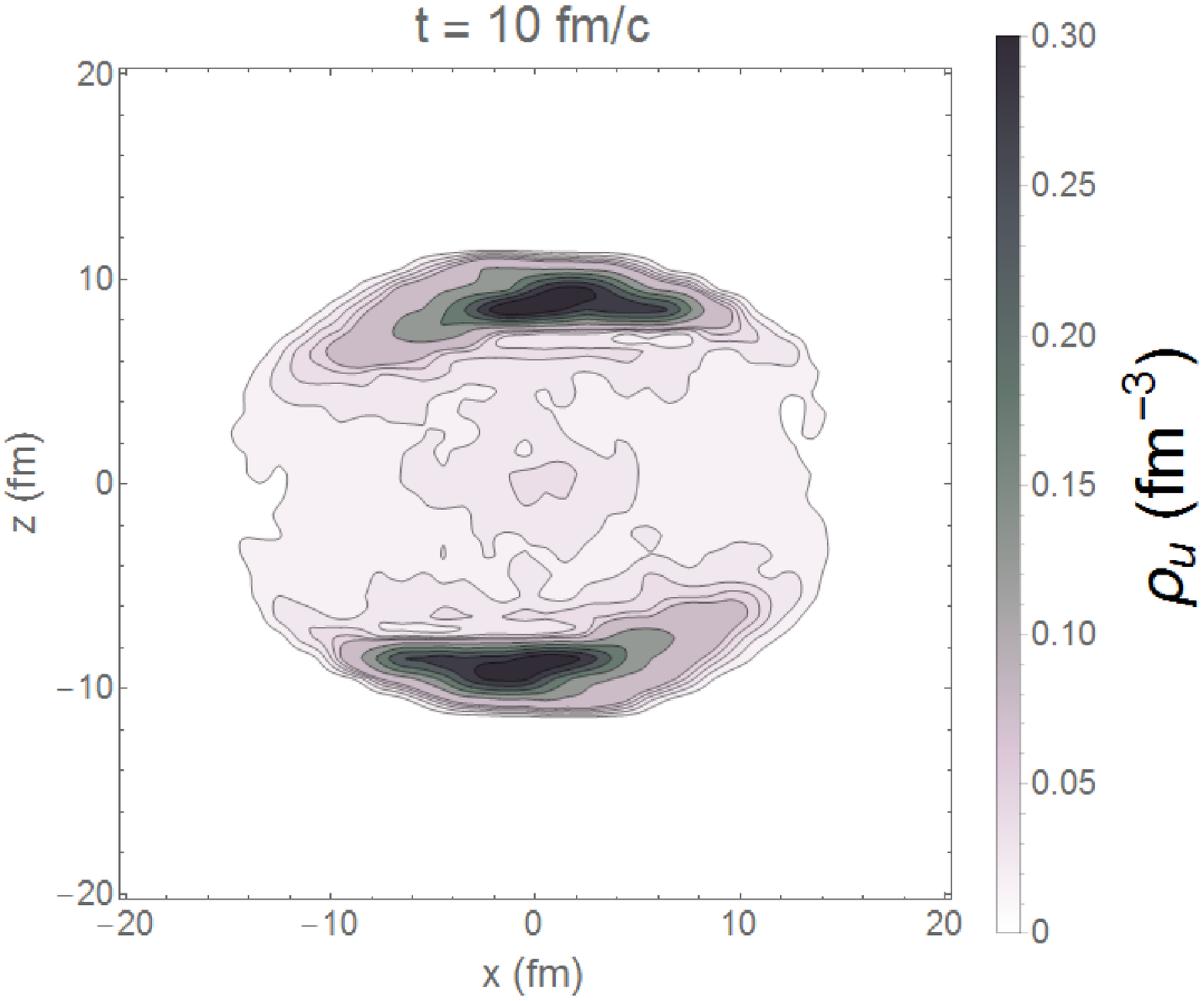}
	\caption{ Density distribution of  an expanding quark matter on the $y=0$ plane at $t=10$ fm/c with (upper window) and without (lower window) a first-order phase transition using the AMPT initial conditions.}
	\label{Gv_0_zpc_s_30_t_10_xz}
\end{figure} 

\begin{figure}[h]
	\vspace{-0cm}
	\centering
	\includegraphics[width=0.45 \textwidth]{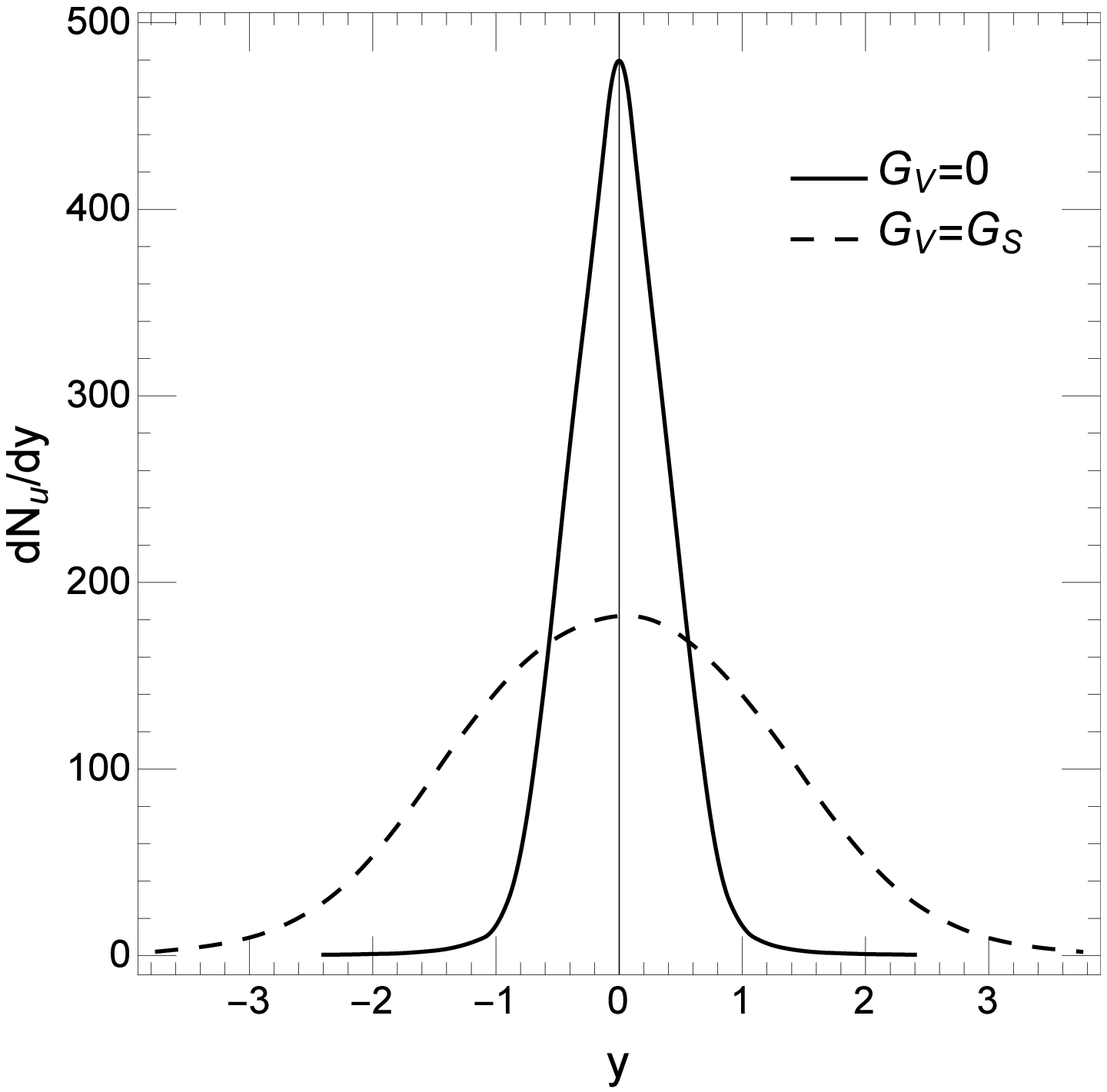}
	\caption{ Final rapidity distribution of  quarks for the cases with (solid curve) and without (dashed curve) a first-order phase transition from an expanding quark matter using the AMPT initial conditions.}
	\label{dNdy_zpc_025}
\end{figure}

Figure~\ref{ini_corr_zpc_25} shows the rapidity and longitudinal coordinate correlation of initial partons from a typical AMPT event for central Au+Au collisions at $\sqrt{s_{NN}}=2.5~\mathrm{GeV}$. This correlation can be quantified as follows:
\begin{equation}
r_{yz}\equiv \frac{\sum_i (y_i-\bar y)(z_i-\bar z)}{\sqrt{\sum_i (y_i-\bar y)^2\sum_i(z_i-\bar z)^2}}=0.355.
\end{equation}
This positive correlation indicates that partons initially on the right side of the quark matter are more likely to have momenta pointing to the right or forward direction, while partons initially in the left of the quark matter are more likely to have  momenta pointing to the left or backward dirction. This correlation helps the initially disc-shaped quark matter to expand, leading to a fast decrease of the density in the center of the  quark matter as shown in the upper row of Fig. \ref{3d_zpc_025_free}.  Here, the  quark matter is initially largely confined in a thin disk of thickness less than 0.5 fm.  When it is allowed to free streaming without any interactions, there appear two high density clumps that fly apart in the opposite directions.   This feature becomes less prominent after the inclusion of quark scattering and mean-field potentials but without a phase transition in the  quark matter, i.e., taking $G_V=G_S$, as shown in the middle row of Fig.~\ref{3d_zpc_025_free}. With a first-order phase transition in the quark matter by setting $G_V=0$, the lower row of Fig.~\ref{3d_zpc_025_free} shows that the initial central disk evolves into three disks of dense matter with one in the middle due to the strong attractions that keep some partons from moving away, besides the two forward and backward moving disks. As the  quark matter expands, these disks transform into rings and finally turn into disjointed clumps. Furthermore, the density distribution of the  quark matter in the reaction plane ($y=0$) shown in Fig.~\ref{Gv_0_zpc_s_30_t_10_xz} indicates that the quark matter with a first-order phase transition expands twice as slow as that without a first-order phase transition. 

Because of the non-trivial spatial distribution even in the case of free-streaming quark matter, the scaled density moments are no longer useful quantities to characterize the density fluctuations of an expanding quark matter due to its spinodal instability or a first-order phase transition.  On the other hand, the different density variations along the beam ($z$) axis shown in Fig.~\ref{Gv_0_zpc_s_30_t_10_xz} are expected to affect the parton rapidity distribution. This is because partons in the middle disc, which is present only in the case with a first-order phase transition, have a small rapidity and due to the attractive quark interactions, they attract partons from the other two discs and slow down their expansion in the longitudinal direction, thus restricting their  rapidities to a narrow region around the midrapidity.  As shown by the solid  line in Fig.~\ref{dNdy_zpc_025}, the parton rapidity distribution in the case with a first-order phase transition is indeed much narrower than that in the case without a first-order phase transition, shown by the dashed  line. This effect can be regarded as a possible signal of a first-order phase transition and is worth studying in  experiments.

\begin{figure}[h]
	\vspace{-0cm}
	\centering
	\includegraphics[width=0.45 \textwidth]{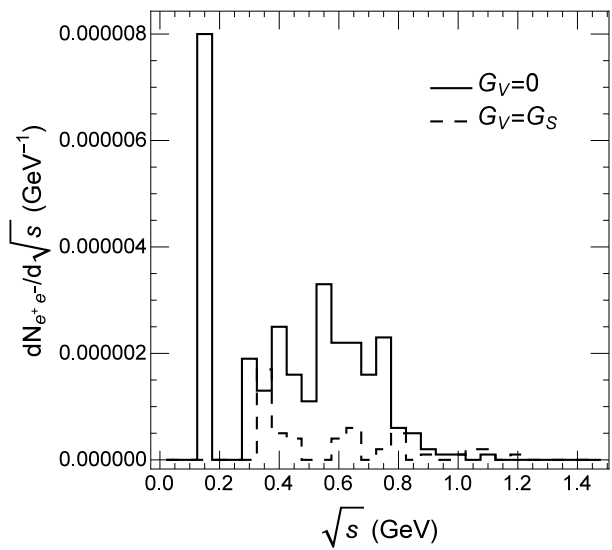}
	\caption{ Dilepton yield as a function of invariant mass $\sqrt{s}$ for the cases with (solid curve) and without (dashed curve) a first-order phase transition from an expanding quark matter using the AMPT initial conditions.}
	\label{electron_zpc}
\end{figure}

We have also studied the dilepton invariant mass spectrum from an expanding quark matter with initial conditions from the AMPT model. This is shown in Fig.~\ref{electron_zpc} by the solid and dashed lines for the cases with and without a first-order phase transition, respectively.  As in the previous section using the blast-wave initial conditions, the presence of a first-order phase transition enhances the dilepton yield as a result of density fluctuations and a longer partonic phase. However, the dilepton yield is lower than  that obtained from the calculation with the blast wave initial condition by two orders of magnitude because there are very few antiquarks in the partonic matter produced in heavy ion collisions at such a low energy and also because we have not included the bremsstrahlung contribution to dilepton production from the quark-quark scattering.

\section{conclusions}

The spinodal instability is a thermodynamic feature of a first-order phase transition in a many-body system. It occurs when its pressure in some parts decreases with increasing density. This  can amplify the density fluctuations and  lead to a phase separation in the system. We have studied this phenomenon by solving the Boltzmann equations using the test particle method. The calculations are based on the NJL model, which has been shown to give good a description of the vacuum properties of the hadrons and also predicts the existence of a first-order phase transition in  
baryon-rich quark matter. We have obtained some intuitive pictures on the phase separation in a quark matter that is either in a static box or undergoes expansion. For the case of a static box, we have found that the growth rates extracted from the early growth of  a sinusoidal density fluctuation agree with the analytical results obtained from the linearized Boltzmann equation. We have also calculated the higher-order density moments  of the quark matter and found them to increase and saturate at large values after phase separation, making them possible signals for the first-order phase transition. The skewness of the quark number event-by-event distribution in a small sub-volume of the  quark matter is also found to increase, but this feature disappears if the sub-volume is large. As for the expanding quark matter, two cases have been studied. One is based on the blast-wave initial conditions, while the other using the AMPT initial conditions, which are disc-like as a result of the strong correlations between the parton rapidity and longitudinal coordinate.  In both cases, we have found that the expansion of the quark matter is slowed down by the presence of a first-order phase transition.  Density clumps are found to appear and lead to an anisotropy in the momentum space, which can be characterized by the scaled density moments and the anisotropic flows $v_2$ and $v_4$, respectively.  An enhancement in the dilepton yield is also observed. The expansion of the quark matter with the AMPT initial conditions is more complex. Normally, the initial disc-like quark matter splits into two discs, moving along the beam axis in opposite directions.  If the  expanding quark matter undergoes a first order-phase transition, a third disc appears in the middle and pulls the other two discs towards it, resulting in a narrower rapidity distribution.

In the future, we  plan to develop a more consistent transport model, in which all  cross sections are calculated self-consistently from the NJL model, so that the temperature and density dependence of the collisional effect can be taken into account. The dilepton production through the $qq\to qqe^+e^-$ process will also be included, since  it could be the main contribution to the dilepton yield from a quark matter of high baryon chemical potential. We also plan to extend the transport model using the PNJL model\cite{Fukushima:2008wg}, which is more realistic and agrees better with the lattice results for a quark matter with low baryon chemical potential.  We hope that our study will help to understand the phase transition in the baryon-rich matter by comparing theoretical predictions with available and future experimental data.

\section*{Acknowledgements}
This work was supported by the US Department of Energy under Contract No. DE-SC0015266 and the Welch Foundation under Grant No. A-1358.

\appendix*

\section{Finite grid size effects}

Counting partons in a grid of finite size in evaluating the mean fields effectively allows the partons in the grid interact with each other, thus modifying  the contact interactions in the NJL model to finite-range ones. To study this effect, we need to calculate the probability for two partons in the same grid to have a separation $\Delta\mathbf x$.  Given a parton located at $x\in[0,a]$ in a 1-dimensional grid $[0,a]$, the probability to find another parton located at $x+\Delta x$ in the same grid is
\begin{eqnarray}
\label{Pdx}
P(\Delta x)&=&\frac 1 a\int dx \theta(x)\theta(a-x)\theta(x+\Delta x)\theta(a-x-\Delta x)\nonumber\\
&=&\mathrm{tri}\left(\frac{\Delta x}{a}\right),
\end{eqnarray}
where
\begin{equation}
\mathrm{tri}(x)\stackrel{\Delta}{=}\mathrm{max}(0,1-|x|).
\end{equation}
The above expression can be straightforwardly generalized to the 3-dimensional case to give
\begin{equation}
\label{Pdx3}
P(\Delta \mathbf x)=\prod_i\mathrm{tri}\left(\frac{\Delta x^i}{a^i}\right),
\end{equation}
where $\{a^1,a^2,a^3\}$ are the grid lengths. The interaction between two partons at $\mathbf x$ and $\mathbf y$ is then replaced by
\begin{eqnarray}
\label{newGs}
G_S\delta^3(\mathbf x-\mathbf y)&\to&\frac{G_S}{\prod_i a^i}\prod_i\mathrm{tri}\left(\frac{x^i-y^i}{a_i}\right),\nonumber\\
\label{newK}
K\delta^3(\mathbf x-\mathbf y)&\to&\frac{K}{\prod_i a^i}\prod_i\mathrm{tri}\left(\frac{x^i-y^i}{a_i}\right).
\end{eqnarray}
Transforming Eq. (\ref{newGs}) from $\mathbf x$-space to $\mathbf k$-space gives
\begin{eqnarray}
\label{newGs2}
G_S\to\tilde G_S&=&G_S\prod_i \frac{2\cos(a^i k_i)-2}{a^i k_i},\nonumber\\
\label{newK2}
K\to\tilde K&=&K\prod_i \frac{2\cos(a^i k_i)-2}{a^i k_i}.
\end{eqnarray}
Note that in the limit that $a^i k_i\to0$ for all the $i$, $\tilde G_S\to G_S$ and $\tilde K\to K$, which means the modification does not affect the long wavelength modes. 

Replacing $G_S$ and $K$ in Eq. (35) in Ref.~\cite{PhysRevC.93.035205} with $\tilde G_S$ and $\tilde K$, respectively, and solving the resulting equation, we obtain the modified dispersion relation, and they are shown in Fig.~\ref{box_check} for a grid size $a^i=2/3~\mathrm{fm}$ by the solid and dashed lines for the cases with and without the collision term, respectively. As expected, the growth rate $\Gamma_k$ is not much affected in the small $k$ region but is significantly suppressed in the large $k$ region. The finite grid size effect is thus similar to the quantum effect shown in Ref.~\cite{PhysRevC.93.035205}. Using a finite grid size essentially allows partons to interact at finite separation, resulting in an effective finite-range interaction.

\bibliography{references}

\end{document}